\documentclass[twocolumn, superscriptaddress, showpacs, amsfonts]{revtex4-1}
\usepackage{amsmath}
\usepackage{latexsym}
\usepackage{amssymb}
\usepackage{graphicx}
\usepackage{hyperref}
\usepackage{array}
\usepackage{color}
\usepackage[update,prepend]{epstopdf}
\newcolumntype{C}[1]{>{\centering\arraybackslash}m{#1}}
\def\ba{\begin{eqnarray}}
\def\ea{\end{eqnarray}}
\def\be{\begin{equation}}
\def\ee{\end{equation}}

\def\bm{\begin{math}}
\def\me{\end{math}}

\def\la{\langle}
\def\ra{\rangle}

\begin{document}
\title{Kinetics of Domain Growth and Aging in a Two-Dimensional Off-lattice System}
\author{Jiarul Midya}
\affiliation{Institute of Physics, Johannes Gutenberg University Mainz, Staudingerweg 7, 55128 Mainz, Germany}
\affiliation{Theoretical Sciences Unit, Jawaharlal Nehru Centre for Advanced Scientific Research, Jakkur P.O., Bangalore 560064, India}
\author{Subir K. Das}
\email{das@jncasr.ac.in}
\affiliation{Theoretical Sciences Unit, Jawaharlal Nehru Centre for Advanced Scientific Research, Jakkur P.O., Bangalore 560064, India}
\affiliation{School of Advanced Materials, Jawaharlal Nehru Centre for Advanced Scientific Research, Jakkur P.O., Bangalore 560064, India}
\date{\today}

\begin{abstract}
We have used molecular dynamics simulations for a comprehensive study of phase separation in a two-dimensional single component off-lattice model where particles interact through the 
Lennard-Jones potential. Via state-of-the-art methods we have analyzed simulation data on structure, growth and aging for nonequilibrium evolutions in the model. 
These data were obtained following quenches of well-equilibrated homogeneous configurations, with density close to the critical value, to various temperatures inside the miscibility gap, 
having vapor-``liquid'' as well as vapor-``solid'' coexistence. For the vapor-liquid phase separation we observe that $\ell$, the average domain length, grows with time ($t$) as $t^{1/2}$, 
a behavior that has connection with hydrodynamics. At low enough temperature, a sharp crossover of this time dependence 
to a much slower, temperature dependent, growth is identified within the time scale of our simulations, implying ``solid''-like final state of the high density phase. 
This crossover is, interestingly, accompanied by strong differences in domain morphology and other structural aspects between the two situations. For aging, we have presented results for 
the order-parameter autocorrelation function. This quantity exhibits data-collapse with respect to $\ell/\ell_w$, $\ell$ and $\ell_w$ being the average domain lengths 
at times $t$ and $t_w$ ($\leq t$), respectively, the latter being the age of a system. Corresponding scaling function follows a power-law decay: 
$~\sim (\ell/\ell_w)^{-\lambda}$, for $t\gg t_w$. The decay exponent $\lambda$, for the vapor-liquid case, is  accurately estimated 
via the application of an advanced finite-size scaling method. The obtained value is observed to satisfy a bound.
\end{abstract}

\maketitle
\section{Introduction}
In addition to having importance from the basic scientific point of view \cite{AJBrayAdvPhys2002, AOnuki:book:2003, RALJones:book:2003, SPuri:book:2009, SiggiaPRA1979, KBinderPRL1974,KBinderPRB1977, 
KBinderBook1991, BAbouPRL2004, SKDasPRL2006, GGKenningPRL2006, SJMitchellPRL2006, LBerthierPRL2007, KBuciorPRE2008, MJAHoreJCP2010, SMajumderPRE2010, SRoyPRE2012, RWittkowskiNatCommun2014,MSleutelNatCommun2014, 
JJungJCP2016, JMidyaJCP2017, JMidyaPRL2017, SRoyPRE2018}, understanding of kinetics of phase transition is of immense relevance in the industrial context 
\cite{CChenRACAdv2016, BTWorrellNatCommun2018,KNomuraNatCommun2019}, having useful applications in designing of new materials and devices. 
A transition from a single phase homogeneous configuration to a multi-phase coexisting situation occurs via the formation and growth of domains of like 
particles \cite{AOnuki:book:2003, RALJones:book:2003, SPuri:book:2009, AJBrayAdvPhys2002}. Depending upon composition or overall density within the systems, various types of domain pattern can 
emerge during the 
evolution to the new equilibrium \cite{SMajumderEPL2011, SiggiaPRA1979, KBinderPRL1974,JMidyaJCP2017, JMidyaPRL2017, SRoyPRE2012, SPuri:book:2009}. When a system, say,
for a vapor-liquid transition, is quenched inside the miscibility gap, with overall (particle) density ($\rho$) close to $\rho_{\rm c}$, the latter being the critical density, 
phase separation proceeds via spinodal decomposition \cite{SPuri:book:2009, AJBrayAdvPhys2002, SMajumderEPL2011, JMidyaPRE2015}. In this case, the morphology consists 
of interconnected, percolating domains. For an off-critical quench, with $\rho\ll\rho_c$ or $\rho\gg\rho_c$, on the other hand, the phase separation occurs via nucleation 
and growth of disconnected droplets \cite{SiggiaPRA1979,KBinderPRL1974,JMidyaJCP2017, SRoyPRE2012, SRoySoftMatter2013} or bubbles \cite{HWatanabeJCP2014, HWatanabeJCP2016, MMatsumotoFluidDynRes2008}.  

The domain patterns are typically characterized via the two-point equal time order-parameter correlation function, $C(r, t)$, defined as \cite{AJBrayAdvPhys2002, SPuri:book:2009} ($r=|\vec{r}|$)
\begin{eqnarray}\label{Eq-corr_fn}
 C(r,t) = \la \phi(\vec{0},t) \phi(\vec{r}, t)\ra - \la \phi(\vec{0},t)\ra \la \phi(\vec{r},t)\ra ,
\end{eqnarray}
where $\phi(\vec{r},t)$ is an appropriately chosen space ($\vec{r}$) and time ($t$)-dependent order parameter. Here the angular brackets represent statistical averaging, 
that may include simulation runs over independent initial configurations. The evolving patterns during phase transitions are typically self-similar in nature \cite{AJBrayAdvPhys2002, SPuri:book:2009}. 
As a consequence, $C(r, t)$ and its Fourier transformation, $S(k, t)$ ($k$ being the wave vector), the structure factor, typically follow the simple scaling 
rules \cite{AJBrayAdvPhys2002, SPuri:book:2009}
\begin{eqnarray}\label{Eq-scale_corr_fn}
 C(r,t) \equiv \tilde{C}(r/\ell), \nonumber
\end{eqnarray}
and 
\begin{eqnarray}\label{Eq-scale_corr_fn}
 S(k, t) \equiv \ell^{-d}\tilde{S}(k \ell). 
\end{eqnarray}
Here $d$ is the space dimensionality and $\ell$ is the average domain length, whereas $\tilde{C}(x)$ and $\tilde{S}(y)$ are two time-independent master functions. 
Such scaling forms point to the possibility of power-law growths of $\ell(t)$ \cite{AJBrayAdvPhys2002, SPuri:book:2009, KBinderPRL1974}, viz.,
\begin{eqnarray}\label{Eq-growth_law}
 \ell(t) \sim t^{\alpha},
\end{eqnarray}
$\alpha$ being referred to as the growth exponent. 

~Another important aspect associated with nonequilibrium dynamics is the aging phenomena \cite{JMidyaJPCM2014, JMidyaPRE2015, DSFisherPRB1988, SRoySoftMatter2019}. This sub-topic of phase transition is relatively less understood. 
For aging related studies, typically one considers the two-time order-parameter correlation function \cite{DSFisherPRB1988}, $C_{\rm ag}(t,t_w)$, defined as 
\begin{eqnarray}\label{auto_Cr}
 C_{\rm ag}(t,t_w)=\langle \phi (\vec{r}, t)\phi(\vec{r}, t_w)\rangle-\langle\phi(\vec{r}, t)\rangle\langle\phi(\vec{r}, t_w)\rangle,
\end{eqnarray}
where $t$ is the observation time and $t_w$ ($\leq t$) is the waiting time or the age of the system. $C_{\rm ag}(t,t_w)$ should decay to 
zero \cite{JPHansen:book:2008, SPuri:book:2009, DSFisherPRB1988} for $t\gg t_w$. In equilibrium, the decay obeys time-translation-invariance \cite{JPHansen:book:2008}, 
i.e., there is overlap of data from different $t_{w}$ when plotted versus the translated time \cite{JPHansen:book:2008} $t-t_w$. 
However, the situation is different in evolving out-of-equilibrium systems. In this case, aging can be observed  \cite{DSFisherPRB1988} because of the slower decay of 
$C_{\rm ag}(t,t_w)$ with the increase of $t_w$.  

In simple nonequilibrium situations $C_{\rm ag}(t,t_w)$ exhibits scaling with respect to $\ell/\ell_w$ as \cite{JMidyaJPCM2014, JMidyaPRE2015, DSFisherPRB1988, SRoySoftMatter2019, SPaulPRE2017}
\begin{eqnarray}\label{scle_Cr}
 C_{\rm ag}(t, t_w) \sim (\ell/\ell_w)^{-\lambda},
\end{eqnarray}
where $\ell$ and $\ell_w$ are the average domain lengths at times $t$ and $t_w$, respectively. Fisher and Huse (FH) \cite{DSFisherPRB1988}, in the context of spin glasses, put bounds on $\lambda$  as
\begin{eqnarray}\label{bound_FH}
\frac{d}{2} \leq \lambda \leq d.
\end{eqnarray}
For usual ferromagnetic ordering \cite{JMidyaJPCM2014, NVadakkayilJCP2019, CYeungPRE1996, CYeungPRL1988}, i.e., typically in systems exhibiting nonconserved order-parameter dynamics, the exponent $\lambda$ satisfies both the 
bounds of FH \cite{JMidyaJPCM2014}, in various space dimensions. However, for conserved order-parameter dynamics, these bounds appear inaccurate \cite{JMidyaPRE2015, CYeungPRE1996}. 
Later Yeung, Rao and Desai (YRD) proposed a more general lower bound \cite{CYeungPRE1996} for $\lambda$, viz.,  
\begin{eqnarray}\label{YRD_bound}
\lambda \geq \frac {\beta+d}{2}.
\end{eqnarray}
This is expected to be valid for both conserved and nonconserved order-parameter dynamics.  Here, $\beta$ is related to the small $k$ 
power-law behavior of the structure factor \cite{CYeungPRL1988}, viz., $S(k, t) \sim k^{\beta}$. For {\it standard} Ising-like ferromagnetic ordering \cite{JMidyaJPCM2014, CYeungPRE1996, CYeungPRE1996, CYeungPRL1988} $\beta=0$. 
Thus, the YRD bound coincides with the lower bound of FH. For conserved dynamics, on the other hand, non-zero values of $\beta$ make the YRD bound different from 
the FH lower bound \cite{CYeungPRE1996}.

The value of $\alpha$ depends upon various features \cite{AJBrayAdvPhys2002, SPuri:book:2009, KBinderPRL1974, SMajumderEPL2011, SMajumderPRE2010, JMidyaJCP2017, JMidyaPRE2015, SRoyPRE2012, SRoySoftMatter2013}, 
e.g., conservation of order parameter, space dimensionality, number of components of the order parameter, range of interaction among the particles, etc. 
In the case of phase separation in solid mixtures \cite{SPuri:book:2009, AJBrayAdvPhys2002}, 
there exists a reasonably unique value, viz., $\alpha=1/3$, referred to as the Lifshitz-Slyozov exponent \cite{AJBrayAdvPhys2002, SPuri:book:2009, IMLifshitzJPCS1961}, implying that 
$\alpha$ is rather insensitive to the choice of system dimension and composition.
In contrast, the process in fluids is much more complex, due to the presence of hydrodynamics \cite{PCHohenbergRevModPhys1977}. There, for critical composition or density, 
$\ell(t)$ cannot be described by a single growth law \cite{AJBrayAdvPhys2002, SPuri:book:2009, SMajumderEPL2011}. The exponents are strongly dependent upon the proximity 
of the final state point to the coexistence curve \cite{SMajumderEPL2011, SRoySoftMatter2013, SRoyPRE2012, JMidyaJCP2017, KBinderPRL1974}. 
In fluids, $\alpha$ shows strong dependence on the space dimensionality as well \cite{HFurukawaPRA1984}. The situation with $\lambda$ is even more complex and largely unexplored. 
Despite being an important aspect, we observe that aging is very poorly understood, except for a few simple situations. 

In $d=3$, understanding of kinetics, including aging, in single- as well as in multi-component systems is relatively more satisfactory, as far as the effects of hydrodynamics are concerned
\cite{AJBrayAdvPhys2002, SPuri:book:2009, SRoyPRE2012, SMajumderEPL2011}. This is more true for critical quenches of the systems 
from a high temperature homogeneous state to temperatures above the triple point \cite{SRoyPRE2012, SMajumderEPL2011}. 
In $d=2$ the status is not even as clean, particularly for off-lattice models. With respect to the effects of hydrodynamics \cite{PCHohenbergRevModPhys1977}, 
in this dimension, there is no general consensus 
\cite{HFurukawaPRE2000, HFurukawaPRA1985, AJWagnerEPL2001, HFurukawaPRA1984, HFurukawaAdvPhys1985, ASinghSoftMatter2015, MSanMiguelPRA1985, RYamamotoPRB1994}. 
While there exists few studies of hydrodynamic domain growth in off-lattice models, in $d=2$, these are primarily for binary mixtures. More importantly, 
aging for two dimensional off-lattice models with hydrodynamics, to the best of our knowledge, has never been studied before.

With the objective of understanding hydrodynamic effects on various aspects of kinetics of phase separation in $d=2$, we consider a generic off-lattice model 
in this paper and perform molecular dynamics (MD) simulations. 
The systems are quenched from a high temperature ($T\gg T_{\rm c}$) homogeneous state to  different sub-critical temperatures, keeping $\rho$ close to $\rho_{\rm c}$. 
For a higher temperature quench, we observe that the domain growth occurs following a power-law: $\ell \sim t^{1/2}$. 
For lower temperatures, on the other hand, a two-step growth is observed: at early time $\ell(t)$ increases by following the same power law 
as the one at high temperature and then at very late time a slower growth of $\ell(t)$ is identified. 
This crossover is due to the fact that the high-density phase becomes ``solid''-like, at late time, for low enough temperature, for which hydrodynamic 
effects are inconsequential. Interestingly, this crossover, at least for the lowest considered temperature, is connected to an unexpected change in the morphological aspect.
This is an important observation coming as a by-product. In addition, we have quantified the scaling behavior for aging at high temperature. In connection with the latter, 
results are also presented on the validity of the YRD bound. These are obtained via the application of an advanced finite-size scaling 
technique \cite{JMidyaJCP2017, SRoySoftMatter2019, NVadakkayilJCP2019}. The role of structure on the overall dynamics has been discussed following the 
calculations of relevant morphological quantities.

The rest of the paper is organized as follows. Section II describes the model and basic methods. In section III we have presented the results, 
alongside the description of the  nonequilibrium scaling method. Finally, the paper has been concluded, with a brief summary, in section IV. 

\section{Model and Methods}
~We consider a model where the interaction between a pair of particles, a distance $r$ apart, is decided by the truncated, shifted and force-corrected 
Lennard-Jones (LJ) potential \cite{MPAllen:book:1987}:
\begin{eqnarray}
V(r)&=&U(r)-U(r_{\rm c}) - (r-r_{\rm c})\frac{{\rm d}U}{{\rm d} r}\Big|_{r=r_{\rm c}};~ \text{for}~ r\leq r_{\rm c},\nonumber \\
  &=& 0;~ \text{for} ~r > r_{\rm c}.
  \label{modyfied_LJ}
\end{eqnarray}
Here  $U(r) \left(=4\varepsilon\left[\left(\sigma/r\right)^{12} - \left(\sigma/r\right)^{6}\right]\right)$ is the standard LJ potential \cite{MPAllen:book:1987, DFrenkel:book:2002} and $r_{\rm c}$ ($=2.5\sigma$) is the cut-off distance, with $\varepsilon$ and $\sigma$ being, respectively, the strength of the potential and the effective diameter of the particles. In our simulations, 
we set both $\sigma$ and $\varepsilon$ to unity. 

~We have performed MD simulations in constant NVT ensemble \cite{DFrenkel:book:2002, MPAllen:book:1987}, by controlling the temperature of the system via a hydrodynamics 
preserving Nos\'{e}-Hoover thermostat \cite{DFrenkel:book:2002, SNoseJCP1984}. We have used square boxes with edge length $L=1024$. Periodic boundary conditions 
are applied along both $x$- and $y$- directions. We set the integration time step for the MD equations at the value $\Delta t= 0.005 \tau$, $\tau$ ($=\sqrt{m \sigma^2/k_{\rm B}T}$) 
being the LJ unit of time. Here $m$ is the mass of a LJ particle and $k_{\rm B}$ is the Boltzmann constant, both being set to unity, for the sake of convenience. 

Initial configurations have been prepared by using random positions and velocities of the  particles,  mimicking $T \gg T_{\rm c}$ situation, with overall particle density $\rho=0.35$. 
These are quenched to different sub-critical temperatures ($T<T_{\rm c}$), where there exists coexistence of vapor phase with ``liquid'' or ``solid''-like phases. 
Note that the values of $T_{\rm c}$ and $\rho_{\rm c}$ for this model are $\simeq 0.41$ and $\simeq 0.37$, respectively \cite{JMidyaJCP2017}.

~For the calculation of $C(r, t$), $S(k,t)$ and $C_{\rm ag}(t, t_w)$, we map our $2D$ off-lattice configurations to square lattices by assigning the value $+1$ 
to a site around which the particle density is higher than an assigned number and $-1$ otherwise  \cite{SMajumderEPL2011, SRoySoftMatter2013}. We have calculated $\ell$ as the distance at 
which $C(r, t)$ crosses zero for the first time. We have obtained $\ell$ also from the first moment of the domain size distribution function \cite{SMajumderEPL2011, SMajumderPRE2010}, 
$P(\ell_i, t)$, viz.,
\begin{equation}
 \ell = \int \ell_i P(\ell_i, t) {\rm d}\ell_i,
 \label{prob_dist_l}
\end{equation}
where $\ell_i$ is the distance between two successive interfaces along any direction, at a given time $t$. 

For the above mentioned calculations we have used the pure domain morphology, i.e., thermal noise from each of the configurations was removed via a renormalization 
procedure \cite{SMajumderPRE2010, SMajumderEPL2011}. The qualitative behavior of $\ell$ obtained from both the described methods are very similar. Thus, throughout the paper 
we have used the numbers from Eq. (\ref{prob_dist_l}). The noise removal procedure was not applied for the calculation of $C_{\rm ag}(t, t_w)$. This is by considering that 
the behavior of the latter may depend on the microscopic details of the systems.  

\section{Results}
\subsection{Quench to a High temperature}
\begin{figure}[ht!]%
    \centering
    {\includegraphics[width=8.5cm]{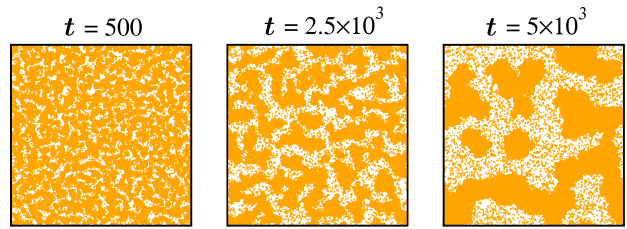}}%
    \caption{Snapshots during the evolution of the two-dimensional Lennard-Jones system, following a quench of a high temperature ($T \gg T_{\rm c}$) 
    homogeneous configuration to the temperature $T=0.375 ~(<T_{\rm c})$, with overall particle density $\rho = 0.35$. The dots represent the locations of the particles.
    }
    \label{Fig:snapshots_T0.375}%
\end{figure}

~Figure \ref{Fig:snapshots_T0.375} shows a set of evolution snapshots taken during the vapor-``liquid'' phase separation in the considered Lennard-Jones model. 
The system is quenched from a high temperature ($T>>T_{\rm c}$) homogeneous state to the temperature $T=0.375$. Formation and growth of domains, comprising of regions 
of high and low densities of particles, can easily be recognized. 

\begin{figure}[ht!]%
    \centering
    {{\includegraphics[width=8.0cm]{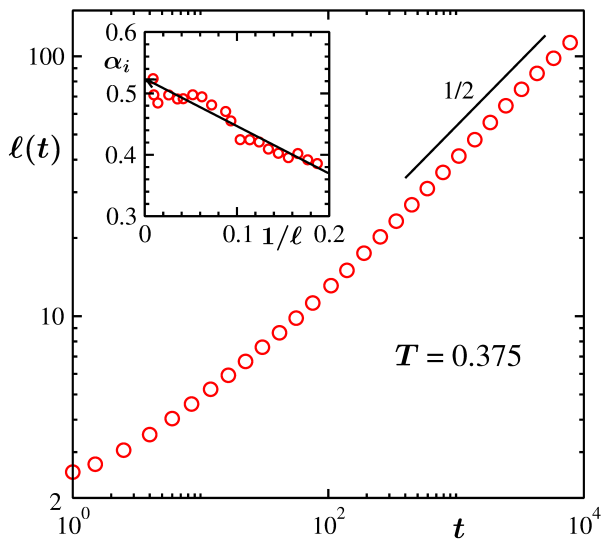}}}%
    \caption{Average domain length, $\ell(t)$, is plotted as a function of time. The solid line corresponds to a power-law with exponent $1/2$. The inset shows the plot of $\alpha_i$, an
    instantaneous exponent, versus $1/\ell$. The arrow-headed line there is a linear guide to the eyes. These results are for coarsening at $T=0.375$.
    }%
    \label{Fig:domainLength_T0.375}%
\end{figure}

~In Fig. \ref{Fig:domainLength_T0.375}, we show the plot of $\ell$, as a function of time, 
on a log-log scale. It appears that there exists a power-law behavior, at least at late time. This may as well be true for the early time data. Considering that there exists a nonzero 
initial length, $\ell_0$, we can write
\begin{eqnarray}
 \ell(t) = \ell_0 + At^{\alpha},
 \label{Eq:dLeng_powerLaw}
\end{eqnarray}
where $A$ is an amplitude. Note that $\ell_0$ is analogous to the background contributions in the context of critical phenomena \cite{JMidyaJCP2017_1, SRoyEPL2011, JVSengersJStatPhys2009}. 
The consistency of the late time data, in Fig. \ref{Fig:domainLength_T0.375}, with the solid line indicates $\alpha \simeq 1/2$. 
This is in agreement with a previous observation for a $2$D phase separating fluid in presence of hydrodynamics \cite{HFurukawaPRE2000}. However, the presence of $\ell_0$, 
even if the latter is of the order of unity, may lead to improper conclusion, if it is drawn by observing the appearance of a data set on a double-log scale. 
For arriving at a better conclusion on the value of $\alpha$, there has been a practice of computing the instantaneous growth exponent \cite{SMajumderPRE2010, DAHusePRB1986, JGAmarPRB1988}, 
$\alpha_i={{\rm d} \ln \ell}/{\rm d \ln t}$. The plot of $\alpha_i$, as a function of $1/\ell$, for the present case, is shown in the inset of the 
Fig. \ref{Fig:domainLength_T0.375}. The data exhibit linear behavior and corresponding extrapolation to the limit $\ell = \infty$ is indeed consistent with $\alpha = 1/2$. 
This plot also states that this exponent is realized from very early time. This is because, for the form in Eq. (\ref{Eq:dLeng_powerLaw}), one expects \cite{SMajumderPRE2010, JGAmarPRB1988}
\begin{equation}
 \alpha_i = \alpha \left[ 1-\frac{\ell_0}{\ell}\right].
 \label{Eq:ins_exp}
\end{equation}
Thus, the early time deviation, that is observed on the double-log scale, is an artefact of the nonzero value of $\ell_0$. For proper validation of the above discussion one 
requires the slope in the inset to match with $-\alpha \ell_0$, the former being $\simeq -0.8$, as can be clearly judged from the plot. Here note that we have $\ell_0 \simeq 2.2$. 
The small deviation of the slope, that may exist, from $-\alpha \ell_0$, can be due to the fact that at very early time there is expected to be a diffusive growth with $\alpha=1/3$, 
however brief the period may be. In absence of the above analysis one would have easily identified a crossover time between $\alpha=1/3$ and $1/2$ as $\simeq 100$.

\begin{figure}[ht!]%
    \centering
    {{\includegraphics[width=8cm]{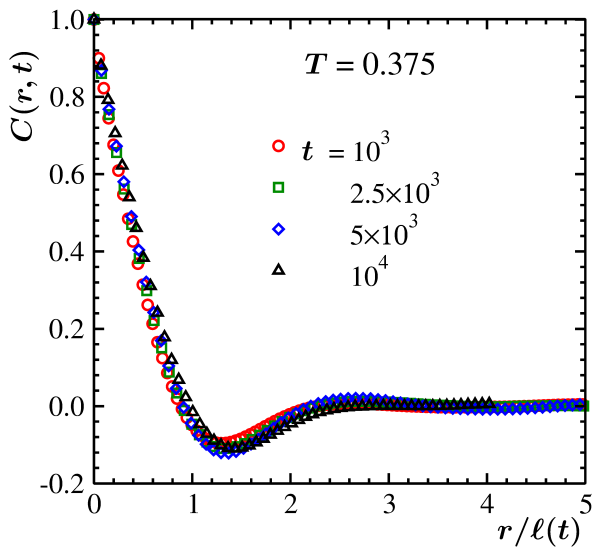} }}%
    \caption{Plots of the two-point equal time correlation function, $C(r, t)$, as a function of $r/\ell(t)$, for $T=0.375$. Data from several times are included.
    }%
    \label{Fig:scorrFun_T0.375}%
\end{figure}

~To characterize the domain morphology, we calculated $C(r, t)$. In Fig.\ref{Fig:scorrFun_T0.375} we have shown plots of $C(r, t)$ from multiple times that are mentioned in the figure. 
Nice collapse of the data from different times, upon scaling of the distance axis by $\ell$, confirms the self-similarity of the growing pattern. This firmly validates the quantitative discussion of 
the growth picture above. The oscillatory behavior of $C(r, t)$ is expected for dynamics for which the system integrated order parameter remains conserved \cite{SPuri:book:2009}. 

\begin{figure}[ht!]%
    \centering
    {{\includegraphics[width=8.5cm]{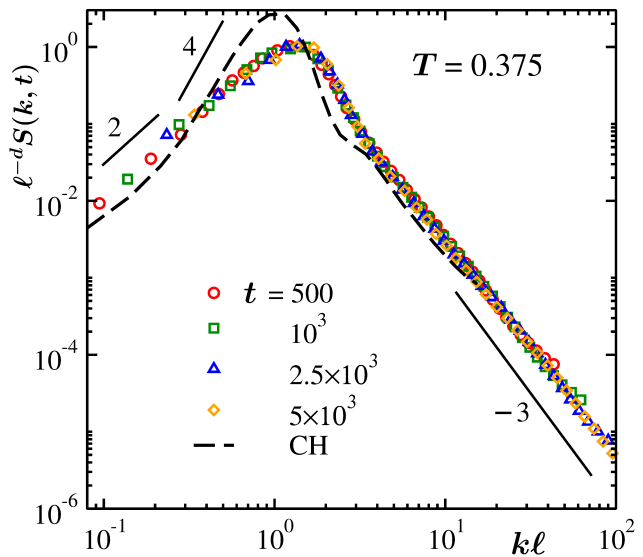} }}%
    \caption{Scaling plot of structure factor: $\ell^{-d} S(k, t)$ is plotted as a function of $k\ell$, on a log-log scale. Data from several times are included. The dashed line 
    corresponds to the scaling plot for the $2$D Cahn-Hilliard (CH) equation. The continuous lines represent different power-law behavior of $S(k,t)$ 
    in various limits or ranges of $k$. Results from MD simulations are for $T=0.375$, whereas the CH data correspond to $T=0.6T_{\rm c}$.
    }%
    \label{Fig:sstFrac_T0.375}%
\end{figure}

In Fig. \ref{Fig:sstFrac_T0.375} we show the scaling plots of the structure factor. There $\ell^{-d}S(k\ell,t)$ is plotted versus $k\ell$. Again, nice collapse of the data 
sets from different times conveys the message that the growth is self-similar. In the small $k$ limit, the power-law enhancement of $S(k,t)$ provides $\beta\simeq 2$, 
whereas the large $k$ behavior follows the Porod-law \cite{AJBrayAdvPhys2002, SPuri:book:2009}, $S(k,t) \sim k^{-3}$, a result of scattering from sharp interfaces in $d=2$ for scalar order parameter.

In Fig. \ref{Fig:sstFrac_T0.375} we have also presented data from the numerical solution of the Cahn-Hillard (CH) equation \cite{SPuri:book:2009, JMidyaPRE2015}. For the latter the set-up is 
made in such a way that the dynamics mimics phase separation in solid binary mixtures with critical composition. Over a part in the small $k$ limit, the CH data are consistent with $\beta\simeq 4$. 
This is in agreement with the prediction of Yeung \cite{CYeungPRL1988} 
and is different from the value obtained for the model that is of our interest here. The tails of $S(k, t)$ in both the cases, however, are consistent with each other. 
Note that in both the cases the phase separation is related to spinodal decomposition and the order-parameter is conserved throughout the process. 
Nevertheless, discrepancy exists, source of which may lie in the difference in the size effects as well as in the composition of up and down spins. 
Rest of the results are from the LJ model.

\begin{figure}[ht!]%
    \centering
    {\includegraphics[width=8.5cm]{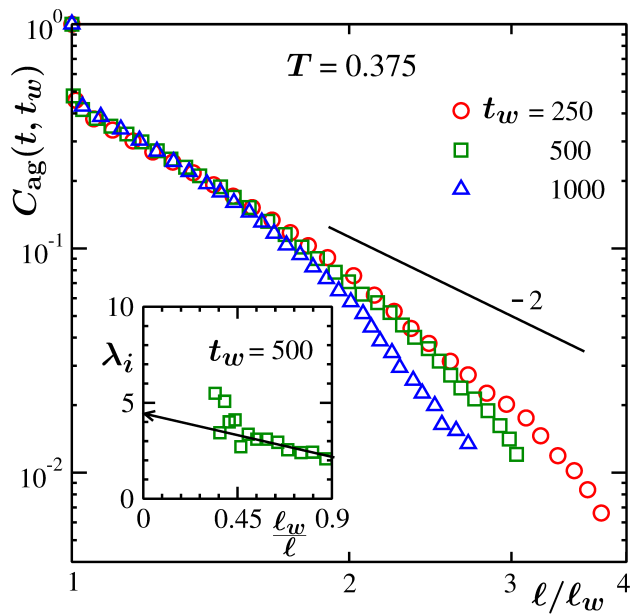} }%
    \vskip 0.1cm
    \caption{Plots of the two-time order-parameter correlation function, $C_{\rm ag}(t, t_w)$, as a function of $\ell/\ell_w$, for a few different values of $t_w$. 
    The solid line represents a power-law with exponent $-2$. The inset shows the plot of the  instantaneous exponent, $\lambda_i$, as a function of $\ell_w/\ell$.
    The arrow-headed line there is a guide to the eyes. These results correspond to the quench temperature $T=0.375$.
    }%
    \label{Fig:sAutocorrFn_T0.375}%
\end{figure}

~Next, we focus on the the aging dynamics. In Fig. \ref{Fig:sAutocorrFn_T0.375} we have shown the scaling plots of $C_{\rm ag}(t,t_w)$, as a function of $\ell/\ell_w$. 
Good collapse of data is observed for smaller abscissa values. There exist continuous bending in the data sets. This is consistent with the results from $d=3$ that 
were presented in Ref. \cite{SRoySoftMatter2019}. The solid line represents the lower bound of YRD, corresponding to $\beta=2$ and $d=2$. 
The bound is certainly satisfied but the actual value of $\lambda$ seems to be much higher than it. It is clear from the plot that for large values of $\ell/\ell_{\rm w}$ 
there is lack of collapse and the data sets with higher values of $t_{w}$ deviate from the mater curve earlier. This is due to the finite-size effects \cite{JMidyaPRE2015}. 
The bending and the size effects make it difficult to arrive at a conclusion on the value of $\lambda$ via simple-minded analysis.

~To get an estimate of $\lambda$, it may be beneficial to calculate the instantaneous exponent $\lambda_i$ \cite{JMidyaPRE2015, JMidyaJPCM2014, DSFisherPRB1988}, defined as
\begin{equation}\label{ins_exp2}
\lambda_i=\frac {{\rm d} \ln[1/C_{\rm ag}(t,t_w)]}{{\rm d} \ln t}.
\end{equation}
A plot of $\lambda_i$, as a function of $1/x$ ($x=\ell/\ell_w$), is presented in the inset of Fig. \ref{Fig:sAutocorrFn_T0.375}. The linear appearance of the finite-size unaffected 
part of the data set there allows us 
to write \cite{JMidyaJPCM2014}:
\begin{equation}\label{lam_lin}
\lambda_i=\lambda-\frac{A_c}{x},
\end{equation}
where $A_c$ is a positive constant. Combining Eqs. (\ref{ins_exp2}) and (\ref{lam_lin}), one obtains an empirical full form \cite{JMidyaJPCM2014} for $C_{\rm ag}(t,t_w)$:
\begin{equation}\label{imp_form}
C_{\rm ag}(t,t_w)=Bx^{-\lambda}\exp\Big(-\frac{A_c}{x}\Big),
\end{equation}
where $B$ is another positive constant. Eq. (\ref{imp_form}) converges to a power-law in the limit $x \rightarrow \infty$, being consistent with 
the theory \cite{DSFisherPRB1988}. 

~One straight-forward way to estimate $\lambda$ is to extrapolate $\lambda_i$, following Eq. (\ref{lam_lin}), to the limit $1/x\rightarrow 0$. The arrow-headed solid 
line in the inset of Fig. \ref{Fig:sAutocorrFn_T0.375} represents such an exercise. The outcome of this exercise suggests $\lambda \simeq 4.8$. For more concrete conclusion, 
below we adopt a finite-size scaling (FSS) analysis \cite{MEFisher:book:1987, MEFisherPRL1972}. This is by considering that the analysis of $C_{\rm ag}(t, t_w)$ is always 
difficult due to noisy nature of data, in addition to the presence of bending and size effects.  

~In the FSS analysis, we introduce a scaling function \cite{JMidyaJPCM2014, JMidyaPRE2015} $Y$: 
\begin{equation}\label{Sca_fun}
Y= C_{\rm ag}(t,t_w) x^{\lambda} \exp(A_c/x).
\end{equation}
This should be independent of the system size, requiring the scaling variable $y$ ($=L/\ell$) to be a dimensionless quantity \cite{JMidyaJPCM2014, JMidyaPRE2015, MEFisherPRL1972, MEFisher:book:1987}. 
Thus, a master curve of $Y$, as a function of $y$, will result from the collapse of data from different system sizes, for appropriate choices of the parameters $A_c$ and $\lambda$. 

Performing simulations for different system sizes, a standard practice for FSS analysis, is always computationally expensive. To reduce such effort we 
re-interpret \cite{NVadakkayilJCP2019, SRoySoftMatter2019} the method here in such a way that the FSS analysis can be performed by taking data, for same system size, 
from different values of $t_w$. This is because of the fact that for different choices of $t_w$, a system has different effective lengths to grow for. 

We express $C_{\rm ag}(t, t_w)$ in terms of $y$ as \cite{SRoySoftMatter2019, NVadakkayilJCP2019}
\begin{equation}\label{new_form_autocorr} 
C_{\rm ag}(t,t_w)=B \exp(- A_c y/y_w) \left(\frac{y_w}{y}\right)^{-\lambda},
\end{equation}
where $y_w = L/\ell_w$. Then, one can rewrite the scaling function $Y$ as \cite{SRoySoftMatter2019, NVadakkayilJCP2019}
\begin{equation}\label{Sca_fun} 
Y(y)= y_w^{\lambda} \exp( A_c y/y_w)  C_{\rm ag}(t, t_w),
\end{equation}
by absorbing $y^{\lambda}$ inside $Y$. In the thermodynamic limit, i.e., for $y\rightarrow \infty$ ($\ell\ll L$), $Y$ is expected to have the power-law behavior: 
\begin{equation}\label{Sca_fun}
Y(y)\sim y^{\lambda},
\end{equation}
which is consistent with the form of $C_{\rm ag}(t, t_w)$ in Eq. (\ref{new_form_autocorr}). 

\begin{figure}[ht!]%
    \centering
    {{\includegraphics[width=8cm]{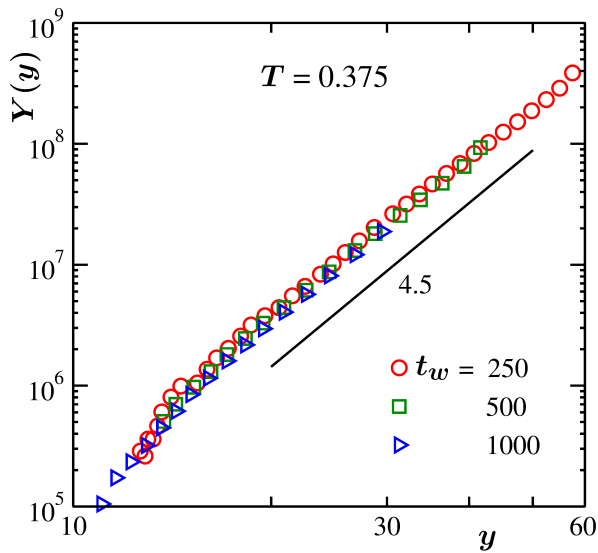} }}%
    \vskip 0.1cm
    \caption{The scaling function, $Y$, is plotted versus the scaling variable $y$. Data from a few different values of $t_w$ have been used. The solid line represents a power-law 
    corresponding to $\lambda=4.5$. These results are for $T=0.375$.
    }%
    \label{Fig:fss-Autocorr_T0.375}%
\end{figure}

~In Fig. \ref{Fig:fss-Autocorr_T0.375} we have plotted $Y$ as a function of $y$, for $L=1024$, by including data from different choices of $t_w$. The best collapse is obtained for 
$\lambda \simeq 4.5$ and $A_c \simeq 2.0$. There the solid line corresponds to a power-law with an exponent $4.5$, which is {\it expectedly} consistent with the master curve. 
The deviation of the master curve from the power-law behavior, at small values of $y$, is related to the finite-size effects \cite{SRoySoftMatter2019, NVadakkayilJCP2019}. 
This estimated value of $\lambda$ is close to the one obtained from the extrapolation of $\lambda_i$. Note that both the values of $\lambda$ 
satisfy the lower-bound of YRD, which is $2$ in this case (recall that we have $\beta \simeq 2$).  

\subsection{Quench to a Low temperature}
~To investigate the effects of temperature on the kinetics we have also quenched the systems to lower temperatures. In this subsection we primarily present results from one of those, 
viz., $T=0.25$, which is significantly deep inside the miscibility gap. Results from other low values of $T$ will be briefly touched upon towards the end. 

\begin{figure}[ht!]%
    \centering
    {\includegraphics[width=8cm]{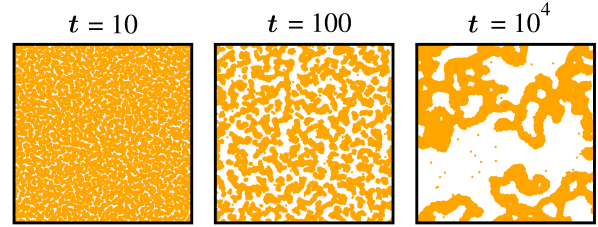} }%
    \vskip 0.1cm
    \caption{Similar to Fig. \ref{Fig:snapshots_T0.375}. But here the snapshots are for quenches to $T=0.25$. 
    }%
    \label{Fig:snapshots_T0.25}%
\end{figure}

~The evolution snapshots from different times, for a representative run, are presented in Fig. \ref{Fig:snapshots_T0.25}. 
The appearance of nearly percolating high-particle-density domains, in the background of low density vapor phase, suggests that phase separation is occurring via spinodal decomposition. 
At early time, the domain morphology is similar to the one observed for the high temperature quench (see Fig. \ref{Fig:snapshots_T0.375}). However, at late time the domains, for the present 
temperature, are spaghetti-like (see the last frame in Fig. \ref{Fig:snapshots_T0.25}). This, as we will see, is due to change in the high density phase with the progress of time. 
Such a change also alters the growth. Because of this, data for any particular regime is not extended over very long period for the chosen simulation run length. This prevents us from 
meaningfully presenting results for the aging phenomena at this temperature.

\begin{figure}[ht!]%
    \centering
    {\includegraphics[width=8cm]{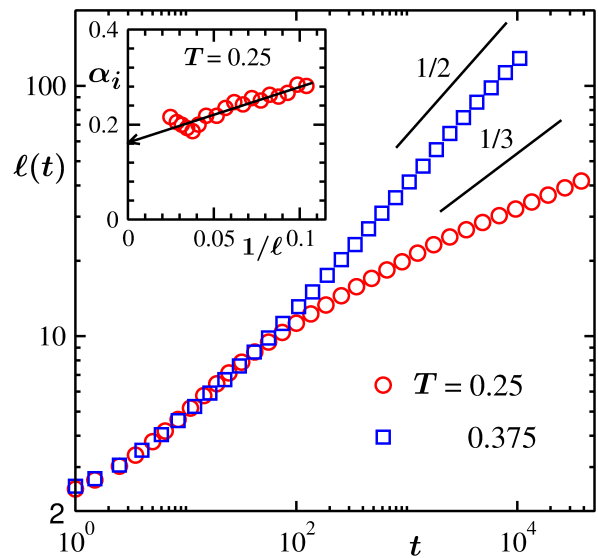}}%
    \vskip 0.1cm
    \caption{Plots of $\ell(t)$ versus $t$ for two different final temperatures, viz., $T=0.25$ and $0.375$. The solid lines correspond to power-laws with exponents $1/2$ and $1/3$. 
    The inset shows the plot of the instantaneous exponent, $\alpha_i$, as a function of $1/\ell$. Here we have shown only the late-time data for the lower temperature quench. 
    }%
    \label{Fig:comb-domainLength}%
\end{figure}

It is clear from Fig. \ref{Fig:comb-domainLength}, where we have shown $\ell$ versus $t$ plots from two temperatures, viz., $T=0.25$ and $0.375$, 
that at early-time ($t \le 100$) the behavior of $\ell(t)$, for $T=0.25$, is very much consistent with the data for high-temperature quench, 
i.e., with $\ell \sim t^{1/2}$. However, at late time, the behavior for $T=0.25$ deviates, the system exhibiting slower growth. 
To estimate the latter, we have calculated $\alpha_i$, for $t>100$. This quantity is plotted as a function of $1/\ell$ in the inset of Fig. \ref{Fig:comb-domainLength}. 
The extrapolation of $\alpha_i$, in the limit $\ell \rightarrow \infty$, provides a value much smaller than even $1/3$. This observation is striking. The value $\alpha=1/3$ 
is typically observed during phase separation in solid mixtures, where growth of domains occurs via particle diffusion. This is well demonstrated via the studies of Ising model 
\cite{SMajumderPRE2010, DAHusePRB1986, JGAmarPRB1988}. For very low temperature quench, however, values much smaller than $1/3$ were also reported \cite{SMajumderPRE2018} for conserved Ising dynamics.

\begin{figure}[ht!]%
    \centering
    {\includegraphics[width=8cm]{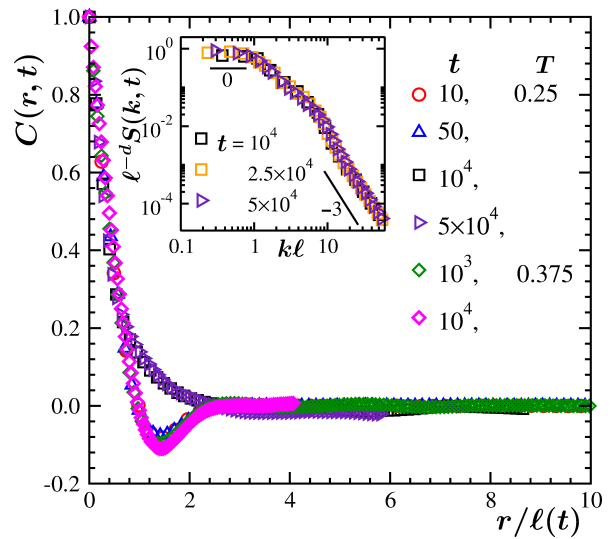}}%
    \vskip 0.1cm
    \caption{Plots of $C(r, t)$ as a function of $r/\ell(t)$, for two considered final temperatures. Data from multiple times, for each of the temperatures, have been presented. 
    Inset shows $\ell^{-d} S(k, t)$ versus $k\ell$ plots, on a log-log scale. The solid lines represent power-laws. Here we have shown only late time results for $T=0.25$. 
    }%
    \label{Fig:comb-scorrFn}%
\end{figure}

In Fig. \ref{Fig:comb-scorrFn} results for $C(r, t)$  are presented. We observe two separate scaling regimes. For comparison, in the same graph we have also presented data from the high 
temperature quench. We find that the early-time behavior of $C(r, t)$ at $T=0.25$ nicely matches with the high temperature data. However, at late time, for $T=0.25$, a crossover has 
happened in morphological feature also. 

\begin{figure}[ht!]%
    \centering
    {\includegraphics[width=8cm]{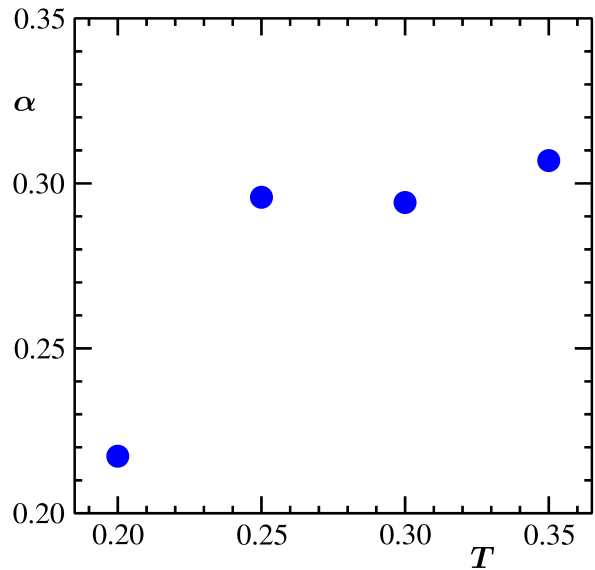}}%
    \vskip 0.1cm
    \caption{Post-crossover regime $\alpha$ is plotted as a function of temperature.  
    }%
    \label{Fig:alphaVsTemp}%
\end{figure}

For low-temperature quench, in the beginning of phase separation, the high density domains are in liquid phase. Thus, the growth of $\ell$ mainly occurs via the hydrodynamics mediated 
flow of materials through the interconnected domains, which leads to the power-law $\ell \sim t^{1/2}$. As the temperature is very low, gradual ordering in particle 
arrangement starts inside the domains, making the effects of hydrodynamics less important at late time. When the domains become ``solid'' like, the growth of $\ell(t)$ occurs via the diffusion of 
particles, which is the growth mechanism in solid mixtures. This explains why the growth is much slower. However, the asymptotic value appears much 
smaller than even the usual picture for solid mixtures. Given that the exponent appears similar to what is obtained for conserved Ising model for very low $T$, we present results for $\alpha$, 
in the late time regime, as a function of $T$, in Fig. \ref{Fig:alphaVsTemp}. With the increase of $T$, clearly the value of $\alpha$ is getting closer to $1/3$. In Ising model, however, the 
morphology does not depend upon $T$ and is similar to fluid phase separation. Interestingly, in the present case the domain morphology is changing with time as the high density 
phase is becoming ``solid'' like. This turns out to be an interesting exception to the understanding that morphology does not depend upon the kinetic mechanism \cite{SPuriPRE1997}. 

~To further characterize the late-time domain morphology, we compute the structure factor. We show $\ell^{-d}S(k, t)$, as a function of $k\ell$, in the inset of Fig. \ref{Fig:comb-scorrFn}, 
from the late times, for $T=0.25$. The scaling of the data from different times indicate the self-similarity of the late-time domains. In the small $k$ limit, flat plateau in $S(k, t)$ 
provides $\beta = 0$, which is different from the high-temperature quench where we found a power-law enhancement with $\beta \simeq 2$. This observation invites fresh studies 
of phase separation in solid mixtures via off-lattice models like the one considered here. However, in the large $k$ limit $S(k,t)$ 
follows the Porod-law quite well.

\begin{figure}[ht!]%
    \centering
    {{\includegraphics[width=8cm]{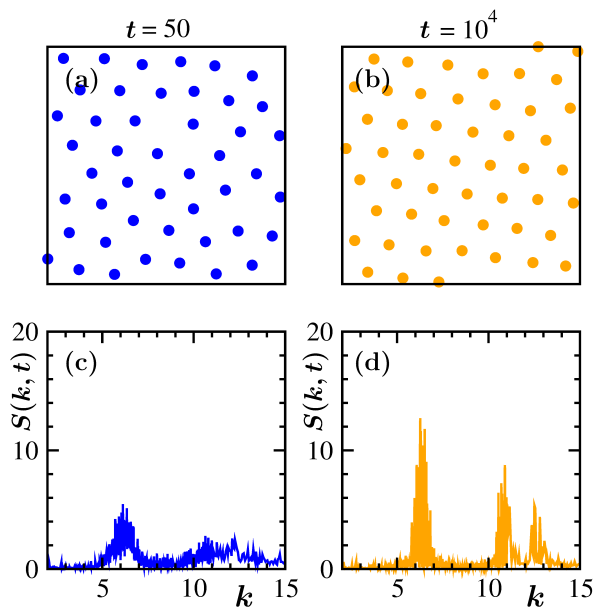} }}%
    \vskip 0.1cm
    \caption{(a) and (b) show enlarged snapshots from two different times for a quench to $T=0.25$. Objective is to depict the arrangement of the particles inside high density 
    regions at early and late times. (c) and (d) show the plots of structure factor, $S(k, t)$, versus $k$, for the snapshot portions in (a) and (b), respectively. 
    These structure factors, $S(k, t) = \sum_{_{i,j}} e^{i \vec{k}\cdot\vec{r}_{ij}}$, $\vec{r}_{ij}=\vec{r}_i-\vec{r}_j$; $\vec{r}_i$ and $\vec{r}_j$ being 
    the positions of the $i^{\rm th}$ and $j^{\rm th}$ particles, were 
    computed by using the off-lattice configurations.
    }%
    \label{Fig:lowTemStructure}%
\end{figure}

~Finally, in Fig. \ref{Fig:lowTemStructure} we have shown parts of domains of high density phase and corresponding structure factors (read caption for definition), $S(k, t)$, 
from the early as well as late times, for growth at $T=0.25$. At early time, the 
arrangement of the particles confirms that the domains are more ``liquid''-like, which gives rise to broad peaks in $S(k)$. On the other hand, at late time, nice hexagonal 
arrangement of the particles provides the evidence of ``solid"-like feature, over intermediate range, which is supported by the sharp peaks in $S(k, t)$. Note that this structure factor 
was calculated by using the off-lattice configurations. 

\section{Summary} 
~We have studied domain growth and aging during phase separation in a model two-dimensional system where the particles interact via the 
Lennard-Jones potential \cite{MPAllen:book:1987, DFrenkel:book:2002}. We have performed molecular dynamics simulations by using the Nos\'{e}-Hoover thermostat that preserves 
hydrodynamics \cite{MPAllen:book:1987, SNoseJCP1984}. For high-temperature quench, we have observed that the average domain size, $\ell$, grows by following a 
power-law $\ell \sim t^{1/2}$, that is related to the hydrodynamic effects. For the low-temperature quenches, at early time, i.e., the time until the domains are in liquid phase, 
the growth of $\ell$ occurs by following the same power-law as the high temperature one. 
However, at late time the growth of $\ell(t)$ drastically deviates from the $t^{1/2}$ behavior and exhibits a slower growth. For very low temperature this exponent is much smaller than even the 
Lifshitz-Slyozov value \cite{IMLifshitzJPCS1961}. Slower growth of $\ell(t)$ occurs due to the late time ``solid''-like feature of the high density domains, which invalidates the hydrodynamics 
mediated flow of materials through the interconnected regions. In this case, the growth of domains primarily occurs via the diffusion of particles. 
But,  a deviation from the Lifshitz-Slyozov like behavior is striking, though there exist an Ising-like temperature dependence of the exponent. This observation requires further attention. 
This is particularly because, change in growth rate is accompanied by a change in domain morphology. 
This is a strong exception to our general understanding that pattern should be independent of kinetic mechanism.

~For high temperature quench, the aging phenomena \cite{DSFisherPRB1988} is also studied. For this purpose we have calculated the two-time order-parameter correlation function \cite{DSFisherPRB1988}, $C_{\rm ag}(t, t_w)$. 
We have shown that the scaling of $C_{\rm ag}(t, t_w)$, with respect to $x$ ($=\ell/\ell_w$), follows power-law decay for large $x$, i.e., $C_{\rm ag}(t,t_w) \sim (\ell/\ell_w)^{-\lambda}$, 
where $\ell$ and $\ell_w$ are the average domain lengths at times $t$ and $t_w$, respectively. The aging exponent $\lambda$ is estimated via the application of 
an advanced finite-size scaling method \cite{JMidyaPRE2015, JMidyaJPCM2014, SRoySoftMatter2019, NVadakkayilJCP2019}. The best data collapse provides $\lambda \simeq 4.5$. 
This estimated value of $\lambda$ satisfies a bound. The number, interestingly, is also in agreement with that obtained for the same model \cite{SRoySoftMatter2019} in $d=3$.

{\bf Acknowledgement:} SKD thanks the Science and Engineering Research Board (SERB) of the Department of Science and Technology, India, for financial support via Grant 
No. MTR/2019/001585. The authors are grateful to K.B. Sinha 
for facilitating an academic travel of JM via his distinguished fellowship grant that was awarded by SERB.

\end{document}